\documentclass{article}
\usepackage{graphicx} 
\usepackage{float} 
\usepackage{authblk}  
\usepackage{anyfontsize}
\usepackage{xcolor}

\title{Recent advances in space sailing missions and technology: review of the 6th International Symposium on Space Sailing (ISSS 2023)}

\author[1]{Elena Ancona}
\author[2]{Roman Ya. Kezerashvili}
\affil[1]{Department of Mechanics, Mathematics \& Management, Politecnico di Bari, 70126 Bari, Italy, \textit{elena.ancona@poliba.it}}
\affil[2]{Physics Department, New York City College of Technology, CUNY, Brooklyn, 11201 NY, USA, \textit{rkezerashvili@citytech.cuny.edu}}

\date{\today}

\begin{document}

\maketitle

\begin{abstract}
The 6th International Symposium on Space Sailing (ISSS 2023) took place on June 5-9, 2023 at the New York City College of Technology, the City University of New York. Since its inauguration in Herrsching (Germany, 2007), the ISSS has been held in New York (USA, 2010), Glasgow (UK, 2013), Kyoto (Japan, 2017) and Aachen (Germany, 2019). 

During the five-day symposium, participants from 14 countries gathered to discuss recent advances in space sailing, investigating new concepts and designs, describing innovative hardware and enabling technologies, strategies for dynamics and control, and providing updates on testing results for systems under development and future mission applications. As part of the 18 sessions, almost 50 oral presentations were held and, subsequently, 17 papers were submitted for review and publication. 

This paper aims to give an overview of all the cutting-edge technologies, detailed analysis and promising results shared with the scientific community as part of the event. Following the noteworthy deployment of the world’s first solar sail IKAROS in 2010, missions like NanoSail-D2 (2011) and LightSail-2 (2019) have showcased the potential of solar sailing technology through successful demonstrations. Besides highlighting advancements in present and future programs, the symposium was an opportunity to reflect on objectives, design and test results from research centers 
and universities, as well as illustrate applications for interstellar travel, evaluate degrading performance and suggest alternative solutions for known limitations. The following Symposium is scheduled for early summer 2025 and will be hosted by TU Delft.    
\end{abstract}

\section{Introduction}

Solar sails, first theorized by K. Tsiolkovsky and F. Tsander back in the 1920s, are large sheets of low-areal density material whose only source of energy is the Sun electromagnetic flux \cite{tsander1969scientific}. At least in theory, a solar sail mission could be of unlimited duration, thanks to the “ever-present gentle push of sunlight” \cite{Vulpetti2008}; a remarkable advantage is that no propellant is needed. Scientists and writers have been dreaming of solar sailing for a long time; interest in solar sails grew in the 1970s with NASA's theoretical studies, though no practical missions materialized at the time. In the following decades, hundreds of theoretical and experimental studies on solar sailing were published. However, researchers in this field lacked regular opportunities to meet and collaborate. This led to the establishment of the 1st International Symposium on Solar Sailing (ISSS), which took place in June 2007 in Herrsching, Germany. The event focused on recent developments in solar sailing technologies and brought together the leading experts in the field. Three years later, in 2010, the 2nd ISSS was held in New York City, USA, where the JAXA team announced the successful deployment of IKAROS, the world’s first interplanetary solar sail (~200 m²), demonstrating spacecraft propulsion using solar radiation pressure (SRP). Advances in solar sail concepts, technologies, dynamics and control, and mission design were later documented in a special issue of Advances in Space Research (ASR) titled "Solar Sailing: Concepts, Technology, and Missions" \cite{KEZERASHVILI20111683}. 
Ten years later, after three more symposia (2013, 2017 and 2019 held in the UK, Japan, and Germany, respectively) and numerous publications in the field, a second special ASR issue reflected the vibrant reality of solar sailing research \cite{KEZERASHVILI20212559}. 
Meanwhile, review papers were also produced regularly on the topic \cite{Fu2016, Gong2019, ZHAO2023}. 

The main goal of this work is to provide an update on recent advances that have been presented at the latest symposium, highlighting achievements and results but also challenges that the community is trying to address. The 6th International Symposium on Space Sailing (ISSS 2023) extended from solar sailing to space sailing and covered a wide range of topics on sails' technology, with an emphasis on innovative mission concepts, materials and testing, control techniques, and mission design. These papers collectively push the boundaries of what is considered feasible with solar (but also magnetic and electric) sails, offering insights into both theoretical innovations and practical applications. Key contributions include the development of advanced mission trajectories for deep space and interstellar exploration, as well as robust control algorithms. Several papers focused on materials research to enhance solar sail durability, while others examined novel control mechanisms for managing momentum and attitude during long-duration missions. The interdisciplinary nature of this research shows how sails are becoming an increasingly viable technology for diverse space exploration missions.

The paper is organized as follows: Section 2 gives an overview of propellantless propulsion systems for space exploration. Section 3 summarizes all achievements is solar sailing until now. In Section 4 details on symposium papers are shared, and conclusions are given in Section 5.

\section{Propellantless systems for space exploration}

Propellantless propulsion systems in space leverage external forces or mechanisms, eliminating the need for spacecraft to carry conventional chemical or electric propellant onboard. An example of a propellantless maneuver is the well-known gravity assist (or "fly-by"), using the pull of celestial bodies to change a spacecraft’s speed and velocity direction. This technique, employed in missions such as Voyager and Cassini \cite{longuski1990automated, cassini}, requires the spacecraft to follow specific trajectories and is analogous to a finite burn, despite not requiring propellant. Other environmental effects can also be used to propel a spacecraft. The primary advantage of these systems is the potential for virtually unlimited mission duration, as they are not constrained by fuel reserves, thus removing the concern of depleting propulsion resources during the mission. This makes them particularly attractive for long-duration space exploration.

The Sun serves as the primary energy source within the Solar System, emitting both electromagnetic radiation and charged particles. Electromagnetic radiation spans the entire spectrum, including ultraviolet (UV), visible, and infrared (IR) light, and carries momentum that can be transferred to objects upon reflection or absorption — this is the principle behind solar sails. Solar sails use the momentum transfer from photons to accelerate spacecraft \cite{Polyakhova,mcinnes1999solar, MatloffBook2005,Vulpetti2008}. Electromagnetic radiation hits the highly reflective surface of the sail, it imparts a small force that continuously pushes the craft forward. Over time, this force accumulates, enabling the spacecraft to reach high velocities. Additionally, the Sun produces the solar wind, a stream of charged particles (mainly protons and electrons) that continuously flows outward. Magnetic sails and electric sails harness the solar wind to generate thrust. Magnetic Sails ("Mag-sails") utilize a large loop of superconducting wire to create a magnetic field that interacts with the solar wind \cite{zubrin1991magnetic, zubrin1991advanced}. Electric sails ("E-sails"), instead, employ long charged tethers that repel solar wind protons (more advantageous than electrons due to their higher mass), to generate thrust \cite{janhunen2004electric, janhunen2010electric}. Comprehensive reviews of the magnetic and electric sails and their application can be found in Refs. \cite{janhunen2010electric,BassettoEL2021,Djojodihardjo2018, BassettoMag2022}. Similar to solar sails but powered by lasers, photon sails use directed energy (from lasers) to accelerate the sail \cite{marx1966interstellar,Lubin2016}. The idea is to build ground-based or space-based laser arrays that emit photons directly onto the sail. This concept is central to the Breakthrough Starshot initiative, which aims to send probes to nearby stars like Alpha Centauri \cite{Breakthrough24}. Some research has also been done on trying to enhance also other radiation sources from stars, artificial sources, or other celestial phenomena \cite{forward1984photon, matloff1988interstellar}.
Besides large area structures like sails, a few other solutions have been proposed, such as electrodynamic tethers, that would interact with the magnetic field \cite{sanmartin1993bare, hoyt2001terminator, landis2007lorentz, forward1976propulsion, TethersLes20}, however,  this technology would be limited in its application by the presence and intensity of the field. 
Each of these methods avoids the need for onboard propellants, relying instead on external forces like sunlight, solar wind, magnetic fields, or gravity. 

When examining forces within the Solar System, solar radiation is significantly more influential in deep space than others. Table \ref{tab:envF} illustrates that both at 1 AU and 5 AU (near Earth and Jupiter, respectively), the effects of solar wind, drag, and magnetic fields are comparatively less dominant \cite{Longuski92, Fu2016}. The provided values correspond to distances of 10 planetary radii from the respective planets. This explains why solar sails have always been considered the most promising propellantless propulsion mechanisms. Nevertheless, researchers have continued to investigate alternatives and possibly hybrid solutions to meet different needs depending on the mission requirements. The ISSS has always welcomed novel approaches and is meant to bring together innovative perspectives.

\begin{table}[h]
\centering
\begin{tabular}{|p{3cm}|p{3cm}|p{3cm}|}
\hline
\textbf{Source} & \textbf{Near Earth} & \textbf{Near Jupiter} \\ \hline
Solar radiation & 9x10$^{-5}$ & 3.3x10$^{-6}$ \\ \hline
Solar wind & 3.1x10$^{-8}$ & 1.1x10$^{-9}$ \\ \hline
Newtonian drag & 7.9x10$^{-11}$ & 5.7x10$^{-7}$ \\ \hline
Magnetic field & 1.9x10$^{-13}$ & 1.6x10$^{-9}$  \\ \hline
\end{tabular}
\caption{Environmental forces in Newton (N)}
\label{tab:envF}
\end{table}

\begin{figure} [h]
    \centering
    \includegraphics[width=0.9\linewidth]{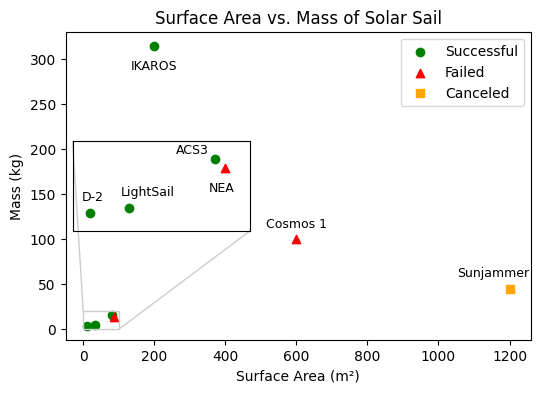}
    \caption{Sail area and mass (including the sail and its associated hardware) for some missions of the past decades.}
    \label{fig:w_A}
\end{figure}

\section{Progress and achievements}
The first attempt to showcase solar sailing was the Cosmos-1 mission, which was a project by Cosmos Studios and The Planetary Society to test a solar sail in space. 
This mission failed due to a launch vehicle malfunction in 2005 \cite{Reichhardt2005} 
Japan's IKAROS mission, launched by JAXA in 2010 \cite{Tsuda2011, KEZERASHVILI20111683, Tsuda2013}, became the first spacecraft to successfully employ solar sail propulsion in interplanetary space. The sail generated measurable thrust, proving that sunlight could effectively power spacecraft for deep space missions \cite{mcinnes1999solar}. In parallel, NASA’s NanoSail-D2 mission \cite{NanoD2011}, launched the same year, validated the viability of solar sailing in Earth orbit on a smaller scale. These early milestones paved the way for more advanced applications. In 2015 LightSail-1 successfully deployed its sail \cite{Biddy2012} and in 2019 LightSail-2 became the first small spacecraft to change its orbit using only sunlight for propulsion \cite{planetary2019lightsail2}. This success of the Planetary Society's sails highlighted their potential for propelling small, fuel-less spacecraft, critical for long-duration missions in deep space. Ongoing research continues to focus on improving sail materials and deployment mechanisms for future interplanetary missions. A recent milestone was achieved with GAMA’s first demonstrator mission, “GAMA Alpha”, launched aboard a SpaceX Falcon 9 on January 3, 2023. The 6U CubeSat, weighing just 12 kg (about the size of a large shoebox), carried a 73 m$^2$ solar sail. The Gama Alpha mission focused on demonstrating successful sail deployment, while its follow-up mission, Gama Beta, will aim to demonstrate solar sail propulsion and navigation \cite{GAMAnews1}. NanoAvionics, a Lithuanian company specializing in nanosatellite platforms, played a crucial role in the development of Gama Alpha, Europe’s first solar sail mission, designed with the support of the French Space Agency, CNES. NanoAvionics provided critical components for Gama Alpha, including electrical systems, telecommunications, and on-board hardware, which were essential in enabling the mission's success \cite{GAMAnews2, GAMAnews3}.
Building on the success of the Gama mission, NanoAvionics also supplied the satellite bus for NASA’s Advanced Composite Solar Sail System (ACS3). ACS3 is a 12U CubeSat (measuring 23x23x34 cm and weighing 16 kg) carrying an 80 m$^2$ solar sail, designed to demonstrate solar sail technology for future small spacecraft applications \cite{Wilkie}. Among its objectives is to showcase solar sails' capabilities for orbit control, including adjusting the semimajor axis to achieve various orbital altitudes. At ISSS 2023, Andres Dono presented ACS3's flight dynamics system, which supports mission planning and solar sail trajectory modeling, and integrates with ground software for orbit determination and real-time analysis. Following its April launch, ACS3 was confirmed as fully operational on August 29, 2024 \cite{ACS3news}.
Table \ref{tab:successful} reports missions that have been successful since IKAROS launch in 2010, whereas failed attempts are listed in Table \ref{tab:failed}. In all circumstances lessons learnt came out of the process: failure has to be considered an option when pushing the boundaries of technology.

\begin{table}[h]
\centering
\begin{tabular}{|c|c|c|p{6cm}|}
\hline
\textbf{Year} & \textbf{Name} & \textbf{Organization} & \textbf{Highlights} \\ \hline
2010 & IKAROS & JAXA & First interplanetary solar sail mission \\ \hline
2011 & NanoSail-D2 & NASA & Solar sail deployed in Earth orbit \\ \hline
2015 & LightSail-1 & Planetary Society & Successful sail deployment in orbit \\ \hline
2019 & LightSail-2 & Planetary Society & First to change orbit using sunlight \\ \hline
2023 & Gama-Alpha & GAMA & 6U CubeSat with 73m$^2$ solar sail \\ \hline
2024 & ACS3 & NASA & 12U CubeSat with 80m$^2$ solar sail \\ \hline
\end{tabular}
\caption{Successful Solar Sail Missions}
\label{tab:successful}
\end{table}

\begin{table}[h]
\centering
\begin{tabular}{|c|c|c|p{6cm}|}
\hline
\textbf{Year} & \textbf{Name} & \textbf{Organization} & \textbf{Highlights} \\ \hline
1999 & Znamya 2.5 & ROSCOSMOS & Reflector failed, broke apart \\ \hline
2005 & Cosmos-1 & Planetary Society & Launch vehicle failure \\ \hline
2008 & NanoSail-D & NASA & Rocket failure prevented deployment \\ \hline
2022 & NEA Scout & NASA & Lost communication after launch \\ \hline
\end{tabular}
\caption{Failed Solar Sail Missions}
\label{tab:failed}
\end{table}

Several solar sail missions have faced significant challenges, resulting in either failure or cancellation. The Cosmos-1 mission, launched by The Planetary Society in 2005, was intended to be the first spacecraft to demonstrate solar sail propulsion. However, it failed due to a malfunction in the launch vehicle, preventing the spacecraft from reaching orbit. NASA's Solar Sail Demonstrator, initiated in 1999, was canceled before launch due to budgetary and technical constraints, stalling further solar sail research at that time. In 2008, NASA’s NanoSail-D mission aimed to test solar sail technology in low Earth orbit, but the Falcon-1 rocket failed, resulting in the loss of the spacecraft.
Russia’s ROSCOSMOS also encountered setbacks with its Znamya 2.5 mission, wide space solar mirror, in 1999. It had a diameter of 25 m, and was expected to produce a bright spot 7 km in diameter, with luminosity between five and ten full moons. However, soon after deployment, the reflector caught on an antenna on the Progress and ripped. The reflector failed to deploy properly and broke apart.

\begin{figure} [h]
    \centering
    \includegraphics[width=0.65\linewidth]{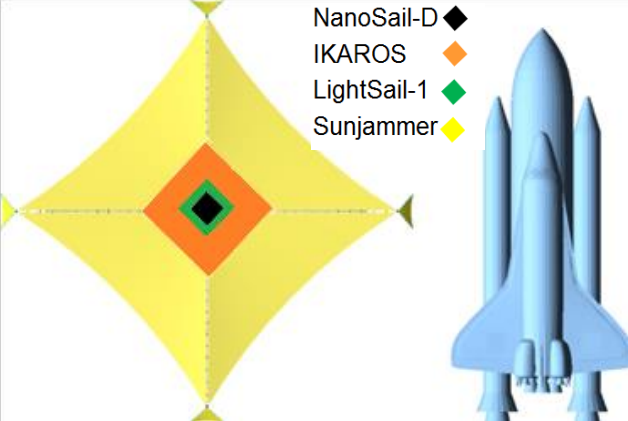}
    \caption{Sunjammer dimensions compared to other sails and the Space Shuttle. Credits O. Eldad and G. Lightsey \cite{SunjammerSmallSat}.}
    \label{fig:sunjammer}
\end{figure}

NASA's Sunjammer mission, led by the private company L’Garde Inc. and planned for 2014, was also canceled due to technical challenges and budget constraints \cite{Sunjammer}. Similarly, the European Space Agency's Gossamer mission was canceled in 2015 after facing technical issues and shifting priorities within ESA \cite{Gossamer11}. Sunjammer, named after a short story written by Sir Arthur C. Clarke, with a surface of 1200 m$^2$ (and a weight of just 45 kg) was meant to be much bigger than sails designed before, as shown in Fig. \ref{fig:sunjammer}. Its area was twice that of Cosmos-1 (600 m$^2$) and around six times that of IKAROS (200 m$^2$).   
More recently, the Near-Earth Asteroid Scout (NEA Scout), a CubeSat mission launched aboard NASA's Artemis I in 2022, was expected to be the first CubeSat to study a near-Earth asteroid using solar sail propulsion. Unfortunately, after launch, communication was never established with the spacecraft, and multiple attempts to deploy its solar sail failed \cite{NEAScout20}. The mission was subsequently declared lost, marking another setback in solar sail development. 
Another interesting concept proposal was the Solar Cruiser, with a surface of more than 1600 m$^2$, which was expected to launch as a rideshare payload alongside the Interstellar Mapping and Acceleration Probe (IMAP) in February 2025 \cite{SolCruiser21}. Although the Solar Cruiser mission was not approved to advance to phase C, its closeout plan included the development and advancement of several key technologies as well as the demonstration of a full quadrant sail deployment, which was successfully completed \cite{SolarCruiserNASA} and achieved NASA’s Technology Readiness Level (TRL) 6 through extensive testing (January 2024) \cite{IAC24_Johnson}.

\begin{figure} [h]
    \centering
    \includegraphics[width=1.0\linewidth]{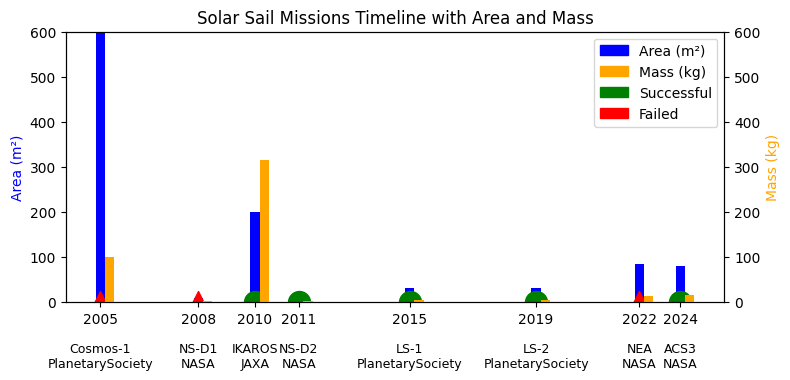}
    \caption{Sail missions timeline and characteristics}
    \label{fig:timeline}
\end{figure}

It is clear that, despite these challenges, valuable lessons from the unsuccessful and canceled missions continue to inform and improve the design and operation of future solar sail projects. Fig. \ref{fig:w_A} shows mass and area features for some of the missions mentioned above. Their characteristics are also reported in Fig. \ref{fig:timeline} and in Table \ref{tab:status_a_m} ordered by their launch dates. Although the basic idea behind solar sailing appears simple, challenging engineering problems have to be solved \cite{LEIPOLD2003, Challenges19}. The Solar Sailing community is actively trying to address these by fostering collaboration and regularly sharing updates on progress in conferences and forums such as the ISSS. Ref. \cite{BERTHET2024101047} presents a recent review and future overlook on space sails for achieving major space exploration goals.

\begin{table}[h]
\centering
\begin{tabular}{|p{2.2cm}|p{1cm}|p{1cm}|p{1.2cm}|p{1.2cm}|p{1.3cm}|}
\hline
\textbf{Mission} & \textbf{Year} & \textbf{Status} & \textbf{M (kg)} & \textbf{A (m²)} & \textbf{D (kg/m²)} \\ \hline
Cosmos-1        & 2005  & F & 100  & 600   & 0.167 \\ \hline
NanoSail-D1     & 2008  & F & 4    & 10    & 0.400 \\ \hline
IKAROS          & 2010  & S & 315  & 200   & 1.575 \\ \hline
NanoSail-D2     & 2011  & S & 4    & 10    & 0.400 \\ \hline
LightSail-1     & 2015  & S & 5    & 32    & 0.156 \\ \hline
LightSail-2     & 2019  & S & 5    & 32    & 0.156 \\ \hline
NEA Scout       & 2022  & F & 14   & 86    & 0.163 \\ \hline
GAMA Alpha      & 2023  & S & 12   & 73    & 0.164 \\ \hline
ACS3            & 2024  & S & 16   & 80    & 0.200 \\ \hline
\end{tabular}
\caption{Solar Sail missions launched, their status (success/fail) and design details: mass, area and resulting areal density.}
\label{tab:status_a_m}
\end{table}

\section{Symposium Overview}
After providing a concise overview of major milestones in space sailing, this section reviews the research presented at ISSS 2023. This review aims to highlight key findings and promising applications, though it cannot fully capture the depth of topics addressed in the five-day symposium. For further details, readers can access the complete set of presentations and papers online \cite{ISSS23}. At the forefront of innovative space sailing concepts, solutions covered various areas: mission design and analysis, advanced materials, attitude control strategies, and deployment methods. The symposium was not only a look toward future missions but also a chance to reflect on past milestones like those related to LightSail-2, ACS3, Gama Alpha and the Solar Cruiser. Multiple presentations referenced these missions, discussing their results and the implications for advancing solar sail research.

\subsection{Mission design and applications}

The Symposium kicked off with key talks by NASA and JAXA on their latest missions. Les Johnson (\textit{NASA Marshall Space Flight Center}) reported on the status of NASA’s Solar Cruiser Solar Sail System and its readiness for heliophysics and deep space missions. With a sailcraft platform using a 2.5-micron thick sail measuring 1,653 m$^2$ with pointing control and attitude stability comparable to traditional platforms, Solar Cruiser's demonstration mission aimed at maturing solar sail propulsion technology and enabling near-term, high-priority breakthrough science missions \cite{LesJ2022}.

\begin{figure} [h]
    \centering
    \includegraphics[width=0.9\linewidth]{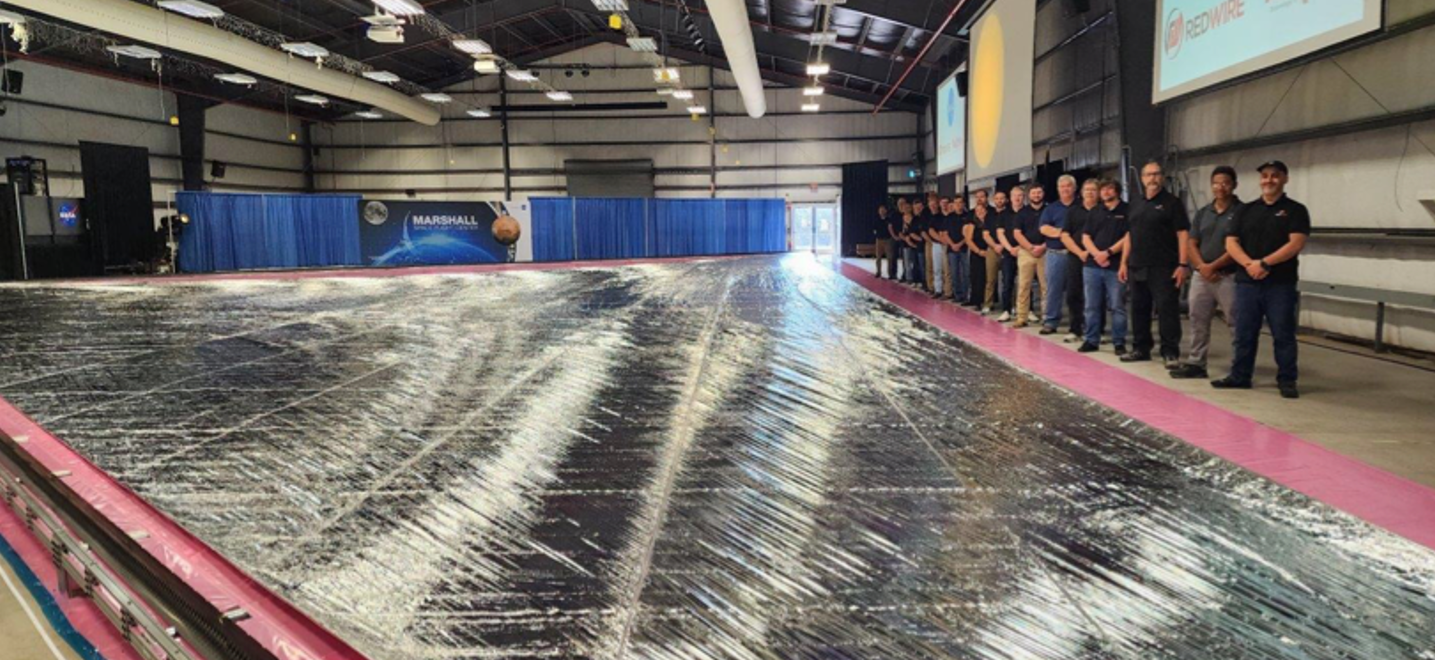}
    \caption{The Solar Cruiser quadrant sail deployment and Solar Cruiser team.}
    \label{fig:team}
\end{figure}

Osamu Mori presented JAXA’s new Solar Power Sail Program, marking a continuation of Japan’s efforts in the Post-OKEANOS Era. OKEANOS (Oversize Kite-craft for Exploration and AstroNautics in the Outer Solar System) is a Japanese mission concept in which electric power is generated by thin-film solar cells integrated with the solar sail membrane \cite{OKEANOS}. The proposed solar sail-powered spacecraft was designed to rendezvous and return from a Trojan asteroid, making it currently the only solution for asteroid sample return. However, due to budget constraints, OKEANOS has not been selected for a JAXA mission. The new solar power sail program aims to expand the technology’s use in various missions by incorporating: i. attitude control for 3-axis stabilization instead of spin stabilization; ii. gimbal-controlled sail orientation to enable simultaneous orbit and attitude control using solar radiation pressure; and iii. enhanced functionality by attaching devices such as an array antenna or interferometer to a boom-type sail.

Andrew Nutter (\textit{GAMA, French space company}) presented the objectives, design, and initial test results of GAMA’s upcoming GAMA-Beta Solar Sail. GAMA’s staged approach to technology development involves a series of missions with increasingly complex sail configurations. Their first demonstrator mission, "GAMA Alpha," successfully deployed a solar sail controlled from a CubeSat, gathering flight data to improve simulations and control algorithms. After two years of rapid development, the Alpha satellite was launched on January 3, 2023, aboard a SpaceX Falcon 9. The next mission, "GAMA Beta," aims to demonstrate controlled navigation in a high Low Earth Orbit (LEO), achieving precise orbit adjustments using photonic pressure alone. Secondary objectives include qualifying systems for deep-space navigation.

The electric solar wind sail (E-sail) \cite{janhunen2004electric, janhunen2010electric} utilizes Coulomb drag propulsion, harnessing solar wind for interplanetary movement. Pekka Janhunen presented the status of E-sail developments and mission designs. Designed for regions beyond Earth's magnetosphere, E-sails use centrifugal force to stretch their tethers. Thrust vectoring is achieved by adjusting the spin plane, and thrust magnitude can be controlled via current and voltage adjustments on the electron gun. For larger spacecraft, multi-tether E-sails are necessary, including auxiliary tethers connecting the main tether tips. In this configuration, differential voltage control allows for independent spin rate and plane control. Multi-tether E-sails can counteract secular spin rate changes due to Coriolis forces, unlike single-tether variants, which require auxiliary propulsion. Coulomb drag propulsion offers a promising approach for orbital debris management and free travel within the solar system.

In its original form, an E-sail consists of a grid of charged tethers held at high positive voltage, interacting with solar wind ions to generate propulsion \cite{janhunen2010electric}. Given the difficulty of deploying large tether structures in deep space, recent work has focused on smaller satellite missions. Lorenzo Niccolai’s (\textit{Pisa University}) study \cite{Niccolai2024} examines deep-space transfers for small spacecraft equipped with E-sails and electric thrusters. Solar panels provide power to the electric thrusters, scaling with the inverse square of heliocentric distance. The thrust model \cite{Mengali2018} describes E-sail propulsion as a function of tether spin plane attitude and Sun-spacecraft distance, solving the optimal control problem using an indirect multiple shooting method based on Pontryagin’s maximum principle.

Many symposium talks covered new mission concepts and examined their feasibility with current or emerging technologies. Interplanetary missions aimed at reaching the heliopause, distant stars, and asteroids were presented.

Observations from 2016 indicate the presence of a small, potentially habitable planet orbiting Proxima Centauri, sparking renewed interest in interstellar exploration \cite{TerrPlanet2016}. At ISSS 2023, several talks examined how solar sails could be employed for interstellar missions.

Bernd Dachwald (\textit{FH Aachen University of Applied Sciences, Germany}) and his collaborators investigated photon-sail capture trajectories within the Alpha Centauri star system, focusing on minimum-time transfer and orbit-raising strategies in habitable zones \cite{Dachwald2004}.

Tim Rotmans (\textit{Delft University of Technology}) studied trajectories from the binary star system towards planet Proxima b \cite{ISSS_Rotmans}. The classical Lagrange points in the binary system are used to find possible trajectories towards Proxima b and heteroclinic connections are exploited using a patched restricted three-body problem method to connect the two phases. Four futuristic sail configurations carrying a payload of 10 grams are used: two sails with a reflective coating on one side and two sails with a reflective coating on both sides. Rotmans work targets the Alpha Centauri system, the destination of the popular Breakthrough Starshot project \cite{Breakthrough24}. Rotmans and colleagues explored graphene-based sails, achieving lightness numbers up to 1779, which drastically reduce travel time. These studies represent a bold step toward interstellar missions powered by light alone, presenting novel trajectory designs that minimize time and energy usage.

Greg Matloff (\textit{City University of New York}) considered an application of Statite-type photon sail probes to achieve rectilinear trajectories to explore the outer solar system and near-interstellar destinations. Statute-type solar photon sails are sufficiently thin and reflective and solar radiation-pressure force on the sail exactly balances solar gravitational force. In such a force-free environment, the spacecraft exits the solar system at its pre-sail-deployment solar-orbital velocity. Here we consider departures from a circular 1-AU solar orbit, the perihelion of a 0.7-1 AU elliptical solar orbit, and the perihelion of a 0.3-a 1U solar orbit. Possible outer-solar-system destinations include Europa, Titan, Enceladus, Arrokoth and more distant such as the Sun’s Gravitational Focus. 

Fil's Project Svarog \cite{ISSS_Fil, Fil2023} discusses a novel mission concept using passively stabilized solar sails to propel spacecraft up to the heliopause. The Svarog system could serve as a low-cost enabler for the testing of new technologies and research opportunities in deep space, piggybacking the increasing number of interplanetary missions, and fostering deep space exploration.

Exploring the outer solar system using solar sail propulsion requires achieving high cruise speeds, which can be facilitated by accelerating the sailcraft in the region near the Sun. Deploying a solar sail close to the Sun maximizes the force exerted by the radiation, enabling higher acceleration and faster cruise speeds. Missions to the Kuiper Belt region and beyond, the Oort Cloud, the gravitational focus of the Sun, and even the Alpha-Centauri system \cite{Lubin2016} are considered the next breakthrough for space exploration. Among various suggested destinations, the trans-Neptunian object Sedna (90377) has recently gained more and more interest from the scientific community \cite{Zubko2021}. Sedna, orbiting the Sun in a highly eccentric orbit, is currently on the way to its perihelion (around 76 AU from the Sun), and recent studies consider this an extraordinary opportunity to get to know more about deep space, being its aphelion at about 936 AU. Elena Ancona (\textit{Politecnico di Bari}) presented the latest results on the ongoing research to reach Sedna with a solar sail \cite{IAC24_Ancona}. The authors recommend using a solar sail that takes advantage of the thermal desorption of its coating as an additional propulsion mechanism, besides its conventional acceleration. During the process of thermal desorption one can neglect the acceleration due to solar radiation because this acceleration is an order of magnitude less than the one due to thermal desorption \cite{Kezer2015, Ancona2019}. It was considered a perihelion of 0.3 AU for and the desorption effects on the cruise speed of the solar sail. Application of this technology with state-of-the-art sails can result in solar-system exit velocities over 100 km/s, as mentioned in previous studies targeting Kuiper Belt Objects \cite{Ancona2019_KB}.


The Mars-Jupiter asteroid belt remains one of the least explored regions of the solar system. The use of a solar sail for a long-term mission into the asteroid belt was considered in Bakhyt Alipova's presentation \cite{ISSS_Alipova}. A complete ballistic calculation of this mission is carried out for a degrading solar sail with a non-ideally reflective surface \cite{Starinova2021}. Authors propose the following ballistic scheme for the mission: i. the launch vehicle takes the solar sail out of the Earth's action sphere; ii. the solar sail makes an Earth-Earth flight, ending with a gravitational maneuver in the Earth's action sphere and entering a trajectory with an aphelion of 3.6 AU and zero inclination; then due to the use of light pressure, a circular orbit is formed with a semi-major axis of 2.9 AU; iii. the solar sail is oriented perpendicular to the direction of the light flux and moves along a twisting spiral. 

Matteo Ceriotti presented the work of a combined study by the University of Glasgow and the Università di Pisa on advances in preliminary solar-sail trajectory design. The proposed techniques for the preliminary design of solar sail trajectories were based on two methods, an artificial neural network (ANNs) \cite{Ceriotti1} and an inverse "shape-based" one \cite{Ceriotti2}. The ANN mimics the structure and learning process of the human brain. “Layers” of neurons, interconnected to each other, are simulated with activation functions, and the outermost layers are fed and provide inputs and outputs of the overall network. The weights of the connections are optimized by “training” the network with known data. Training the ANN on a sub-set of known transfers between near-Earth asteroids, the authors demonstrated that ANNs can estimate the duration of new solar-sail transfers with good accuracy in modest computational time. 
With the shape-based trajectory design method instead, a geometrical curve for the trajectory, connecting a departure and target orbit or state, is established analytically by means of tuning the shape parameters. This allows to obtain an approximate thrust profile. The authors introduced novel shaping functions to describe the 3D time evolution of the spacecraft state vector. The shape optimization problem is solved using a genetic algorithm, in which a set of shape coefficients and the initial and final spacecraft position are computed, while enforcing suitable constraints on the magnitude and direction of the solar radiation acceleration vector. Numerical simulations of transfers to potentially hazardous asteroids demonstrate the feasibility of this method to estimate of solar sail trajectories.

Besides interstellar exploration, other applications were discussed. As the number of space debris in low Earth orbits keeps increasing, new strategies for active removal must be conceived to prevent the overcrowding of this orbital region. In this regard, the use of propellantless propulsion systems, such as solar sails, might enhance the effectiveness of this removal strategy, and could allow a larger orbital element space to be explored, while also extending the mission duration at the same time. Bianchi et al. \cite{ISSS_Bianchi} studied blended locally optimal control laws for space debris removal in LEO using a solar sail. Locally-optimal laws are used to control the transfer to maximize (or minimize) the rate of change of a specific orbital element or a suitable combination of them, depending on the transfer phase. The proposed strategy consists of reaching the orbit of a target debris object, collecting it, and bringing it down to a lower altitude where the aerodynamic forces can eventually lead to its natural decay and burn. When blended control laws are used to target the debris, the optimal weighting factors are calculated with a genetic algorithm to assess the relative importance of each orbital element. The disposal orbit must be carefully chosen to allow the sail to increase its altitude again to target a new object. The dynamical model used in this study considers the effect of solar radiation pressure (including eclipses), aerodynamic forces and Earth's oblateness. The same analysis has been repeated in Ref. \cite{Bianchi2024} for a solar sail with performance characteristics similar to those of NASA’s ACS3 mission.

Patric Seefeldt (\textit{DLR}) presented the High Performance Space Structure Systems GmbH (HPS) family of dragsail systems, ADEO \cite{ADEO21, IAC24_ADEO}. The ADEO subsystem is a scalable drag augmentation device that uses the residual Earth atmosphere present in LEO to passively de-orbit satellites between 1 and 2000 kg. For the de-orbit maneuvre, a large surface is deployed which multiplies the drag-effective surface of the satellite significantly. Thereby the drag force is increased causing accelerated decay in orbit altitude. The drag augmentation device is that it does not require any active steering and can be designed for passive attitude stabilization, thereby making it applicable for non-operational, tumbling spacecrafts as well. First commercial candidate missions to be targeted were already identified by Airbus DS (DE) and QinetiQ (BE).

Solar sails can become a key technology to provide that future with a constant flow of materials to Earth and from it. Rozhkov \& Starinova \cite{ISSS_Rozhkov} proposed using a solar sail to ensure cyclic heliocentric motion of a cargo spacecraft between Earth and other planets. Following \cite{Hughes2006} and applying Pontryagin’s maximum principle they define Hamiltonian and solve the boundary value problem with consideration of non-ideal reflection and degradation. The simulation is carried out for 4 loops of cyclic motion Earth-Mars-Earth and Earth-Mercury-Earth to demonstrate the possibility of the suggested transport system.

The Helianthus concept, developed by Università di Roma la Sapienza and in collaboration with the Italian Space Agency, was presented by Giovanni Vulpetti. The project aims at realizing a sailcraft for a geostorm early-warning with warning times longer than 100 minutes for the solar fast streams. Helianthus should operate on a range of orbits quasi-synchronous with the Earth-Moon Barycenter. This goal could be reached with four maneuvers per year by offsetting the initial orbital errors, the sail wrinkles perturbation, and the Earth-Moon system’s gravitational disturbance.

There are many commonalities between solar reflectors and solar sails. In terms of their structure, the reflector would be susceptible to the same flexible modes as those of a solar sail, as well as wrinkling of the reflective film. The SOLSPACE project proposes the use of large orbiting reflectors in a polar orbit to enhance the energy output of 
farms, particularly at dawn and dusk, when output is low but energy demand is high \cite{Celik2022a}.  The proposed solar reflectors are hexagonal in shape, with a side length of 250 m and a total mass of order 3000 kg \cite{Celik2023}. Iain Moore with collaborators from the University of Glasgow provided an overview of the SOLSPACE project, detailing the commonalities between the proposed solar reflectors and solar sails. He discussed potential synergies between solar reflector and solar sail development, which offers benefits for both systems, and explored applications where the solar reflectors can be used as a solar sail to provide an adaptive platform for secondary missions \cite{Celik2022b}.

\subsection{Hardware development and testing}

Hardware development is essential for the future of solar sail missions, and several papers at ISSS 2023 addressed this. 

Zachary McConnel ({\it{Space Systems, Redwire}}) provided an overview of the results from the test of a full-scale quadrant for the 1,653 m$^2$ Solar Cruiser Sail and outlined critical lessons learned that will inform ongoing efforts to develop the technology towards flight further \cite{ISSS_McConnel}. Concerning the same mission, Maddox reported on the full-scale deployment of NASA’s Solar Cruiser, where they tested the structural integrity of the sail under simulated space conditions \cite{ISSS_Maddox}. This testing ensured that the sail could withstand the harsh space environment, providing critical data on material behavior and mechanical resilience. 

As a result of its many years of experience in deployable masts and their deployment control mechanisms, DLR has developed a new generation of deployment mechanisms for its in-house manufactures Collapsible Tubular Masts (CTMs) presented by Marco Straubel ({\it{DLR Institute of Lightweight Systems}}). These mechanisms are characterized by a modular setup and geometric scalability. This allows for the realization of 4-mast deployment modules for solar sails as well as 2-mast modules for solar arrays or a 1-mast module for magnetometer booms with the very same basic concept.

As of today, all launched solar sails have a square shape, and such design is related to the simplicity of its deployment mechanism. Kezerashvili ({\it{City University of New York}}) presented a comparison of two novel concepts to deploy and stretch the large size circular solar sail based on: i. the superconducting current loop attached to the sail membrane \cite{Kez1,Kez2}; ii. the inflatable toroidal shell attached to the sail membrane \cite{KEZERASHVILI202317}. In the framework of a strict mathematical approach based on the theory of elasticity, elastic properties of a circular solar sail membrane, inflatable toroidal shell, and superconducting wire loop are analyzed. The magnetic field induced by a superconducting wire with engineering current densities achievable today can deploy and stretch the large membrane. The formulas for the superconducting wire and sail membrane stresses and strains caused by the current in the superconducting wire are derived. Numerical calculations for the sail of a radius of 5 m to 150 m attached to a superconducting wire with a cross-section radius of 0.5 mm to 10 mm are performed. By introducing the gas into the inflatable toroidal shell one can deploy and stretch a large size circular solar sail membrane \cite{KEZERASHVILI202317}. The analytical expressions obtained for both types of deployment mechanisms can be applied to a wide range of solar sail sizes. Numerical calculations for the sail of radii up to 100 m ($\sim$ 30,000 m$^2$) made of CP1 membrane are presented.
The authors performed a comparative analysis and demonstrated the feasibility of the deployment and stretching of a solar sail with a large-size circular membrane attached to a superconducting wire loop or the inflatable toroidal shell.

Andrew Heaton from NASA Marshall Space Flight Center in his talk went through the mission timeline for NEA Scout, including several milestones related to the solar sail technology. This was to be the first deep space solar sail deployment by NASA, the first deep space mission with a scientific target for rendezvous, the first solar sail mission to use a moving mass to manage momentum and the first NASA solar sail to use under full 3-axis control. Several hardware challenges have been addressed, including calibrations of the Adjustable Mass Translator, the Reaction Control System, and the sail disturbance torque model.

These contributions illustrate the critical importance of hardware development and testing, and deployment methods development in ensuring the reliability of solar sail technology for future space missions.

\subsection{Control techniques}

Attitude control is a critical aspect of solar sail design, as the orientation of the sail directly affects its thrust direction. Conventional spacecraft attitude control methods, such as control moment gyros, reaction wheels, and thrusters, are generally unsuitable for solar sails, especially thrusters, since solar sails are intended to operate without propellant. Ideally, the sail’s center of mass ($c.m.$) aligns with its center of pressure ($c.p.$), but minor manufacturing and deployment errors can create an offset, causing unwanted torque that must be corrected for stable attitude. Conventional attitude control methods cannot effectively maintain the sail’s attitude for several reasons, as first discussed by Wie \cite{Wie2004}. Fortunately, by having an attitude control strategy that allows a controlled $c.m.$-$c.p.$ offset, desired attitude torques can be generated. Fu et al. \cite{Fu2016} give a comprehensive overview of the most common methods for attitude control, for both rigid sailcraft such as control vane, gimbaled/sliding masses, shifted/tilted wings and non-rigid ones. Non-rigid sailcraft rely on internal tension from centrifugal forces and cannot use mechanisms that apply large out-of-plane loads. For non-rigid sails, an attitude control method using reflectivity-controlled membranes has been developed: by adjusting voltage, the membrane's reflectivity changes, shifting the cp and generating a torque for attitude control.

Precise control is essential for the success of solar sail missions. Given the relevance of the topic, many papers presented at ISSS 2023 focused on attitude control techniques.

Inness et al. \cite{ISSS_Inness} discussed momentum management for the NASA Solar Cruiser, using Active Mass Translators and thrusters to counteract the buildup of momentum caused by solar radiation pressure. The work ensures that the spacecraft maintains stability during long-duration missions by balancing solar radiation and gravitational forces. 

An overview of the blended locally optimal control laws proposed by Bianchi et al. has been already detailed in section 4.1 when presenting their debris removal mission \cite{ISSS_Bianchi}.

A sailcraft in low Earth orbits experience accelerations different from the theoretically predicted ones. In order to quantify these discrepancies between the real and modeled solar-sail dynamics, Livio Carzana et al. \cite{ISSS_Carzana} develop a set of calibration steering laws presented at the ISSS 2023. These steering laws allow us to characterize the acceleration envelope of the sail, that is, they allow to quantify the solar-sail acceleration at every sail orientation and identify the contributions due to different sources of acceleration. In particular solar radiation pressure, planetary radiation pressure, and aerodynamic drag. This paper provides a first definition and preliminary assessment of a range of calibration steering laws. The analyses presented make use of NASA’s ACS3 mission as a baseline scenario and account for different possible orientations of its orbit as well as operational constraints (e.g., limiting the solar sail’s maximum attitude rate of change) \cite{Carzana2024}. The results highlight the benefits and drawbacks of each steering law and the impact that they have on the orbital elements, with a particular focus on the orbital altitude.

Another talk presented by Carzana on behalf of the Delft University of Technology team related to a new model for the planetary radiation pressure acceleration for optical solar sails provided a novel planetary radiation pressure acceleration model for optical solar sails. This model is an extension of the “spherical” planetary radiation pressure acceleration model for ideal solar sails devised in \cite{Carzana2022}. The accuracy of the optical model compared to the ideal model is analyzed and discussed. The effect of planetary radiation pressure on the maneuvering capabilities of solar sails in LEO is investigated.

Solar sails operating in the space environment experience deformations in sail shape that result in relatively large disturbance torques which dictate the required performance of the spacecraft attitude control and momentum management systems. Using as a reference the Solar Cruiser, Gauvain expanded on the consequences of these shape changes \cite{ISSS_Gauvain}. The deformed sail model results demonstrate considerably higher induced disturbance torques compared to the simplified assumption of a flat sail with a $c.m.$-$c.p.$ offset. Accurate prediction of these disturbances is crucial for the spacecraft attitude control system, thus it is recommended to begin medium/high fidelity modeling as early as possible in the design cycle.

Near Earth Asteroid (NEA) Scout was a mission to test solar sail propulsion for orbital transfer from cislunar space to flyby and image an asteroid. One of the goals of the mission was to characterize the solar torque on the sail to ensure successful attitude control for the orbit transfer and imaging the asteroid. Using the generalized model for solar sails \cite{Reyes2005}  in Ref. \cite{Reyes2007}  developed a general process to update the torque tensor coefficients using estimates of sail torque over a range of directions to the sun. Ben Diedrich \cite{ISSS_Diedrich} ({\it{NASA, Marshall Space Flight Center}}) adapted and implemented this process for the specific case of NEA Scout mission. This process met the needs of the NEA Scout mission and can be adapted to characterize the solar torque for other missions with different sails.

Ryan Caverly et al. presented the Cable-Actuated Bio-inspired Lightweight Elastic Solar Sail (CABLESSail) concept that will enable robust, precise, and scalable attitude control of solar sails \cite{ISSS_Caverly}. This concept leverages lightweight cable-driven actuation \cite{Caverly2014} to achieve large, controllable elastic bending and torsional deformations in the booms of a solar sail that mimic the motion of an elephant’s trunk or a starfish’s arms. These large cable-driven boom deformations, which are actuated using winches located near the solar sail’s center of mass, modulate the shape of the entire sail to create an imbalance of solar radiation pressure to induce control torques in all three solar sail axes. This actuation method scales well with an increase in solar sail size, as cables can transmit forces over kilometers in length from a lightweight and small stowed volume. 

Some effort has been made on the operations of a solar sail in close proximity of an asteroid \cite{CeriottiJGCD2016,CeriottiASR2021}. In order to maximize the scientific return of the mission, asteroid close proximity operations will be essential, including hovering. 
Zitong Lin ({\it{University of Glasgow}}) proposed a second-order sliding mode control law for hovering above asteroids using solar sails \cite{ISSS_Zitong}. This robust control strategy is designed to overcome the challenges posed by underactuated sail systems, particularly in proximity operations where precise control is needed for missions like asteroid reconnaissance. Such advances in control techniques are vital for enabling complex maneuvers in space. The authors simulated the case of hovering on a displaced orbit above 433 Eros. 

Orbital motion of solar sails is controlled by steering the orientation of the sail relative to the sun. Using an active shape control method for spinning solar sails Yuki Takao’s study proposes a new deployable payload that has an autonomous flying ability using a solar sail \cite{ Takao2019}. Similarly to the orbiting experiment of deployable payloads in the Hayabusa2 mission, multiple solar sails are inserted in a periodic orbit around the small body \cite{Takao2022}. Each solar sail changes its orbit, forming a constellation around the small body. The constellation in low-altitude orbits makes it possible to perform high-resolution global mapping.

Alesia Herasimenka from the University Cote d’Azur, reported the sufficient and necessary conditions for the controllability of solar sails and proposed a novel necessary condition for local controllability of solar sails in orbit about a celestial body \cite{Alesia1}. This condition inspects whether a non-ideal sail is capable of generating an arbitrary variation of its current orbital elements, i.e., decrease or increase any function of the Keplerian integrals of motion. The authors propose an efficient methodology to verify the aforementioned condition numerically. A suitable convex relaxation is introduced to study the optimality system associated with maximizing the change of the sail orbital parameters in each direction along one orbit. The suggested methodology is then extensively used to determine minimum requirements on the reflectivity of the sail allowing local controllability for any orbital configuration \cite{Alesia2}. The presented results are conservative and independent of the gravitation constant as well as two out of five Keplerian orbital elements and may be used for preliminary mission analysis purposes. The proposed methodology is generalized to tackle station-keeping applications around an arbitrary periodic orbit (e.g., Halo, Lyapunov) \cite{Alesia3}.

The Planetary Society’s LightSail-2 solar sail mission showed the enormous value of cameras to image the sails for engineering assessment and public outreach. LightSail-2 was the first mission to demonstrate controlled solar sailing in a small spacecraft, in this case, a 3U CubeSat \cite{Spencer2021}. Bruce Betts ({\it{The Planetary Society}}) presented images of sail deployment captured dynamics of the process, and allowed for rapid verification of successful deployment enabled an engineering assessment of sail deployment, evolution of the boom and sail shape, and sail degradation with time. The presentation by Justin Mansell complemented Betts' overview on the mission giving the orbit evolution and attitude control performance and correlated changes in the orbit with both On-Off control and solar activity. They discussed the recalibration of the gyros and compared deorbit simulations to mission results to highlight the effect of solar sailing and gain insights relevant to drag sail spacecraft.

\subsection{Materials and technologies}

Advancements in materials science are key to extending the lifespan and performance of solar sails. Carzana’s new optical radiation pressure model incorporates material degradation and varying optical properties \cite{ISSS_Carzana2}, allowing more accurate mission planning for solar sails near Earth and deep space. The study presented a new analytical model for planetary
radiation pressure (PRP) acceleration, in particular for optical solar sails. The results also show that neglecting the PRP acceleration from the dynamics highly affects the results, as in that case the maximum relative errors on the altitude and inclination increases are 12\% and 55\%, respectively.

Rozhkov and Starinova's (\textit{Samara National Research
University}) work on solar sail material degradation focused on how the sails' reflectivity decreases over time, which is a critical factor in long-term missions \cite{ISSS_Rozhkov}. The same effects were also considered as part of Alipova's work \cite{ISSS_Alipova}, where a degradation factor is used to determine the value of the optical characteristic at which the performance of the sail would be drastically reduced. This focus on material resilience is crucial for ensuring that solar sails can withstand the rigors of space over extended periods, improving their viability for future missions.

Benjamin Gauvain (\textit{NASA Marshall Space Flight Center}) illustrated how crucial it is to model properly the sail deformation as early as possible given the high influence on the membrane shape and performances, using as reference the Solar Cruiser \cite{ISSS_Gauvain}.

Within this area of research, the work presented by the authors of this review paper focused on the use of thermal desorption. By coating the solar sail with materials that undergo desorption at elevated temperatures, one can introduce an additional acceleration mechanism \cite{Ancona2019, Ancona2019_KB}. As the sail heats up from solar radiation at a specific heliocentric distance, the coated material desorbs, releasing additional energy that boosts the sail's speed \cite{Kezer2015}. This effect, combined with the traditional solar radiation pressure, provides an extra source of propulsion. The perihelion of the solar sail's orbit is selected based on the temperature required for the desorption of the coating material. At this critical perihelion, the sail, that was stowed, is deployed and reaches its maximum temperature, triggering desorption and subsequent acceleration. The sail achieves escape velocity from this process but continues to accelerate further due to the radiation pressure. By leveraging both the increased solar radiation pressure near the Sun and the additional thrust from material desorption, this approach offers a promising method for reaching high velocities, making it highly suitable for missions to the outer solar system and beyond \cite{KEZERASHVILI20212577}.

Radiation pressure on a non-absorbing body is attributed to optical scattering, e.g., reflection, refraction, or diffraction. To achieve the optimized spiral orbit maneuvers a sun-facing solar sail that uniformly scatters light perpendicular to the sunline is desirable.  Electromagnetic pressure driven sails navigate through space by transferring momentum from natural light sources such as the Sun or engineered sources such as lasers to the sailcraft. Motivated by the potential to reach solar escape velocities or relativistic speeds, laser-driven sails have attracted considerable scientific attention over the past several decades \cite{Marx1966,Forward1984,Beals1988,Friedman1988,Frisbee2009,Lubin2016,Daukantas2017}. Today photonic metamaterial technologies offer the potential to create thin actively controlled diffractive films that provide optomechanical characteristics such as a high momentum transfer efficiency and switching between diffraction orders. A uniform passive diffraction grating provides a force that is independent of the illumination point \cite{Swartzlander2017,Swartzlander2018a,Swartzlander2018b}. In contrast a space variant grating such as a “bigrating” \cite{Swartzlander2019a,Swartzlander2019b} may provide a position-dependent force. Thus, a space-variant grating may be designed to function as a “beam rider”. A laser-driven beam rider must produce self-action both to pull the sail into the beam path when disturbed and to inhibit tumbling, i.e. the stability of a light sail riding on a laser beam \cite{ Manchester2017,Popova2017}.
The theory of radiation pressure on a diffractive solar sail was developed in Ref. \cite{Swartzlander2022}.  Grover Swartzlander reported the recent study performed by the team of scientists and engineers from Rochester Institute of Technology, Johns Hopkins Applied Physics Laboratory, and NASA George C. Marshall Space Flight related to the comparison of diffractive films for solar sailing.  Using an open source finite difference time domain software package called MEEP optimized meta-surface films with optimized prism gratings was compared. 

Due to the higher efficiency and unique maneuverability options diffractive solar sailing can theoretically provide, it can allow spacecraft to attain novel observational orbits such as high solar inclination angles above the ecliptic. Amber Dubill's report “Next in Solar Sail Technology: Diffractive Solar Sailing” presented an expand the original concept of a singular solar polar orbiter by means of diffractive solar sailing, to an entire $4\pi$ steradian constellation which corresponds to the heliophysics community’s interests of multi-view, simultaneous observations of the Sun. Authors explore options for unique diffractive sail optical designs and space environment resistant materials in the scope of efficiency and manufacturability by means of analysis and experimentation. 

The solar sail membranes generally are metalized thin films. During their lifetime the functional surfaces are exposed to various types of corpuscular radiation mainly resulting from solar flares, a solar wind, coronal mass ejections, and solar prominences which include low and high-energy electrons and protons, helium ions, and ions of some light metals \cite{Kezerashvili20071,Kezerashvili20072}. Sznajder et al. \cite{Sznajder2021} investigated the influence of the recombination processes of solar wind protons with metal electrons on the thermo-optical properties of the sail membrane. The results indicated a harsh degradation after only a few days of exposure to the interplanetary solar wind environment depending on the surface temperature. Because of the drastic degradation observed due to proton irradiation, Erik Klein (\textit{DLR and University of Bremen}) and his collaborators investigated the development of the specular reflectance over the accumulated fluence and mapped results of the study to certain mission scenarios, so that an understanding of the process with its change of specular reflectance over mission time is gained. The authors derived the progression of the specular reflectance over fluence and mission time. The analysis allows a more accurate assessment of the performance of solar sails for future missions.

For the ACS3 mission, several candidates of metalized thin film polymer membrane materials and composite laminate materials were evaluated for the solar sail. Simulated space environment tests for many of these materials have been performed using ultraviolet, electron, and proton radiation in ground-based laboratories \cite{Sznajder2021, Kang2021} However, there has been limited information on the lifetime and durability of these materials under the combined effects present in a space environment. During 2019 to 2021, space environmental effects on these materials were evaluated on the International Space Station (ISS). Jin Ho Kang and his collaborators (\textit{NASA, Langley Research Center and Glenn Research Center}) reported the results of this experiment for ACS3 solar sail membrane materials and composite boom laminate samples that were exposed to the space environment outside of the ISS (vacuum UV, atomic oxygen, solar particle events, etc.). The samples were returned to Earth for post-flight analysis. Pre- and post-flight material properties including morphological, optical, and thermal properties, were evaluated to assess the effects of space environmental exposure on the samples.

Besides large reflecting membrane areas as means of propulsion solar sailing missions also need power supply. Tom Sproewitz’s presentation addressed this issue in the project DEAR (Deployable 100 W PV Array for SmallSats): a deployable 100 W solar array is developed that can be stowed in and deployed out of a 1U CubeSat volume. 

All these contributions show the critical importance of degradation effects on solar sail performance and material science research in ensuring the reliability of solar sail technology for future space missions.

\section{Conclusions}
New concepts and mission designs, innovative hardware and enabling technologies, as well as strategies for dynamics and control, were presented to the public during the 6th International Symposium on Space Sailing held in 2023 at the CUNY New York City College of Technology \cite{ISSS23}. The ISSS 2023 offered a rich platform for collaboration, knowledge-sharing, and exploration of the latest developments in solar sailing technology. The symposium underscored the progress made since early missions like IKAROS and LightSail-2, showcasing how advancements in materials, control strategies, and mission design are steadily addressing the engineering challenges of propellant-less propulsion. Recent missions such as NASA’s ACS3 and Solar Cruiser, GAMA-Beta Solar Sail, and post-OKEANOS development were also explored, providing valuable insights into the performance and potential applications of solar sail technology. These projects, along with other innovative research presented, are pivotal in realizing future interstellar and deep space exploration goals. With a strong emphasis on continued innovation and collaboration, the symposium highlighted the importance of evolving this field through international cooperation and shared insights. As the community prepares for the next ISSS at TU Delft in 2025, the discussions and findings from ISSS 2023 will undoubtedly be helpful in the definition of new space exploration concepts.


\end{document}